\documentclass[conference]{IEEEtran}
\ifCLASSINFOpdf
   \usepackage[pdftex]{graphicx}
\else
   \usepackage[dvips]{graphicx}
\fi
\usepackage{amsmath}

\usepackage{stfloats}
\usepackage{amssymb}
\usepackage{color}
\usepackage{subfigure}
\usepackage{graphicx}

\newtheorem{lemma}{\bf{Lemma}}
\newtheorem{theorem}{\bf{Theorem}}
\newtheorem{proposition}{Proposition}

\begin{document}
%
\title{On the Fairness Performance of NOMA-based Wireless Powered Communication Networks}


\author{\IEEEauthorblockN{Yong~Liu\IEEEauthorrefmark{1}\IEEEauthorrefmark{3},
Xuehan~Chen\IEEEauthorrefmark{2}\IEEEauthorrefmark{3}, Lin X.~Cai\IEEEauthorrefmark{3}, Qingchun~Chen\IEEEauthorrefmark{1},
Ruoting~Gong\IEEEauthorrefmark{4},and Dong~Tang\IEEEauthorrefmark{1}}
\IEEEauthorblockA{
\IEEEauthorrefmark{1} School of Mechanical and Electric Engineering, Guangzhou University, Guangzhou, 510006, China\\
\IEEEauthorrefmark{2} School of Software, Central South University, Changsha, 410075, China\\
\IEEEauthorrefmark{3} Department of Electrical and Computer Engineering, Illinois Institute of Technology, Chicago, IL 60616, USA\\
\IEEEauthorrefmark{4} Department of Applied Mathematics, Illinois Institute of Technology, Chicago, IL 60616, USA\\
Email:ly454580194@gmail.com,
xhss1991@163.com,
\{lincai,rgong2\}@iit.edu,
\{qcchen,tangdong\}@gzhu.edu.cn
}
}

\maketitle

\begin{abstract}
The \emph{near-far} problem causes severe throughput unfairness in wireless powered communication networks (WPCN).
In this paper, we exploit non-orthogonal multiple access (NOMA) technology and propose a fairness-aware NOMA-based scheduling scheme to mitigate the \emph{near-far} effect and to enhance the max-min fairness.
Specifically, we sort all users according to their channel conditions and divide them into two groups, the interference group with high channel gains and the non-interference group with low channel gains.
The power station (PS) concurrently transmits energy signals with the data transmissions of the users in the interference group. Thus, the users in the non-interference group can harvest more energy and achieve a higher throughput, while the users in the interference group degrade their performance due to the interfering signals from the PS.
We then apply order statistic theory to analyze the achievable rates of ordered users, based on which all users are appropriately grouped for NOMA transmission to achieve the max-min fairness of the system.
Meanwhile, the optimal number of interfered users that determines the set of users in each group, is derived.
Our simulation results validate the significant improvement of both network fairness and throughput via the fairness-aware NOMA-based scheduling scheme.
\end{abstract}


%
\IEEEpeerreviewmaketitle

\section{Introduction}
Radio frequency (RF) enabled wireless energy transfer (WET) technology provides a promising solution to power wireless devices with continuous and stable energy over the air \cite{lu2015wireless}, \cite{clerckx2018fundamentals}.
In a wireless powered communication network (WPCN) \cite{bi2015wireless}, the power station (PS) transfers energy to wireless users and the access point (AP) collects the data of wireless users, which are referred to as wireless energy transfer (WET) and wireless information transfer (WIT) respectively.

Resource allocation for WPCN has been investigated in several works, with focus on improving the network throughput or energy-efficiency.
In \cite{zhao2016optimal}, the time block was divided into two stages for WET and WIT respectively, and the time allocation was studied such that the network throughput was maximized.
The time allocation and power control were investigated in \cite{wu2016energy} to maximize the energy-efficiency of a WPCN, where the time for WIT was shared by multiple users in time division multiple access (TDMA) manner.
Some recent works have shown that non-orthogonal multiple access (NOMA) technology [5]-[8] 
is an effective method to improve the spectral efficiency.
A NOMA-based scheduling scheme for a WPCN was proposed in \cite{abd2017non}, and the uplink NOMA technology was utilized for WIT, where users simultaneously transmitted their information to the AP.
The transmit power and time were jointly optimized in \cite{chingoska2016resource} to maximize the expected sum-rate with given average power constraints in a NOMA-based WPCN.
The spectral and energy efficiency of TDMA-based and NOMA-based WPCN were compared in \cite{wu2018spectral}.
While, most existing works considered non-orthogonal transmission for information delivery, and only a few works studied how to explore the non-orthogonal transmission for both WET and WIT.
The energy-efficiency of NOMA-based WPCN with concurrent WET and WIT was studied in \cite{zewde2018noma}. However, the interference caused by simultaneous operation of WET and WIT was ignored. In addition, the most existing works on NOMA-based WPCN mainly focused on the network throughput maximization without considering the fairness among users.

Due to the path loss attenuation of RF signals, the user far from the PS harvests a smaller amount of energy, and the user far from the AP needs more energy. This is referred to as \emph{near-far} phenomenon, which leads to a severe fairness issue in a WPCN.
Adaptive time and power allocation schemes were studied  in \cite{ju2014throughput}\cite{diamantoulakis2016wireless} to enhance the minimum rate of all users, namely to achieve the max-min fairness.
Adaptive power and time allocation under three fairness criteria including the max-min, the proportional and harmonic fairness, were compared in \cite{guo2016convexity}.
A proportional fairness maximization problem was formulated in \cite{diamantoulakis2017maximizing} to optimize the transmission time.
In these works, the time was dedicated for either WET or WIT, and the positions of all users were assumed a priori. 
To the best of our knowledge, how to exploit NOMA for both WET and WIT to improve the network fairness under random user deployment is not explored yet, which is the motivation of this paper.

In this paper, we propose a fairness-aware NOMA-based scheduling scheme for a WPCN with random user deployment.
To achieve the max-min fairness, we first sort all users based on their channel conditions, including the channels for WET and WIT. Then, we divide users into two group, the interference group with good channel conditions and the non-interference group with poor channel conditions.
By exploiting NOMA transmission, the PS concurrently transmits energy signals with the data transmissions of the users in the interference group.
Therefore, the achievable rates of the users in the interference group decrease due to the interference from the PS, and those of the other users increase due to more harvested energy.
By applying order statistics theory,  we analyze the achievable rates of users and the achieved max-min fairness of the system. We show that the minimum rate of the interfered users  decreases with respect to the number of interfered users, and that of the non-interfered users monotonically increase. Thus, the optimal number of interfered users  is derived to maximize  the minimum rate of all users, or equivalently, to achieve the max-min fairness.
In addition, as users harvest more energy due to the concurrent WET and WIT, the network throughput is improved as well, which is validated by the simulation results.

The remainder of this paper is organized as follows.
The system model and NOMA-based scheduling scheme for a WPCN are described in Section II.
The fairness-aware NOMA-based scheduling scheme is proposed and analyzed in Section III.
Simulation results are provided in Section IV, followed by concluding remarks in Section V.

\section{System Model and Problem Formulation}

\subsection{System Model}

\begin{figure}
  \centering
  \subfigure[System model of a wireless powered communication network]{
    \label{Fig:Model} 
    \includegraphics[width=3.4in]{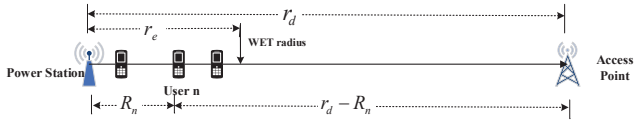}}
  \subfigure[Time structure of NOMA-based scheduling scheme]{
    \label{Fig:Protocol} 
    \includegraphics[width=3.4in]{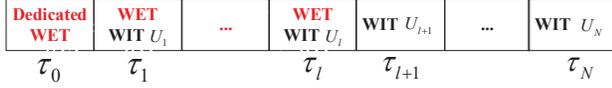}}
  \caption{System model and time structure}
  \label{fig:subfig} 
\end{figure}

We consider a wireless powered communication network (WPCN), which consists of $N$ energy-constrained users  $\mathcal{N}=\{1,2,\cdots, N\}$, one power station (PS), and one information access point (AP) as shown in Fig. \ref{Fig:Model}.
The PS charges all users via transmitting energy signals, and the users utilize the harvested energy to send information to the AP. All devices work in half-duplex mode.
The users are randomly deployed between the PS and the AP.
Denote $r_e$ and $r_d$ as the wireless energy transfer (WET) radius and the distance between the PS and AP, respectively.
We define $R_{n}$ as the distances from the user $n$ to the PS.
Therefore, and the distance from the $n$-th user to the AP is $r_d-R_n$, and $R_n\sim U[0,r_e]$.

Block fading channel is considered such that the channel does not vary during the unit time block but may independently change from one block to another. Let $\frac{L_0d_0^{\alpha}}{R_{n}^\alpha}$ and $g_n(t)$ denote the large and small scale fading coefficients of the link between the PS and the user $n$ in the $t$-th time block, where $L_0$ represents the path loss at reference distance $d_0$ and $\alpha$ is the path loss exponent. Similarly, we define $\frac{L_0d_0^{\alpha}}{(r_d-R_{n})^\alpha}$ and $h_n(t)$ as the large and small scale fading coefficients of the link between the user $n$ and the AP in the $t$-th time block.
The transmission power at the PS is $P_s$. $\sigma^2$ represents the noise variance at all users and the AP.


\subsection{NOMA-based Scheduling scheme for Wireless Powered Communication Network}

A NOMA-based scheduling scheme for WPCN is exploited. As is shown in Fig. \ref{Fig:Protocol}, each time block is slotted, and WET and WIT can be scheduled in the same or different time slots.
The fixed time allocation is considered.
$\tau_0$ is dedicated for WET, and the remaining time slots are used for WIT, where the WET is concurrently operated during the data transmissions of some users.
Let $\tau_n=\frac{1-\tau_0}{N}$ denote the allocated time of the user $n$.
All users are divided into two groups, the interference group $\mathcal{N}_0$ with $|\mathcal{N}_0|=l$ users, and the non-interference group $\mathcal{N}_1$ with $|\mathcal{N}_1|=N-l$ users.
The user grouping policy that determines the users of each group is a very critical issue for network fairness. 
Beside the  dedicated WET time, when the user $i$ in the interference group $\mathcal{N}_0$ transmits information to the AP, the PS simultaneously transfers energy to the other users, thus the total time for WET is $\tau_0+\sum_{i\in\mathcal{N}_0} \tau_i$, which is larger than $\tau_0$ and implies more energy harvested at each user.
If $l=0$, our proposed NOMA-based scheduling scheme is degraded to the conventional TDMA-based scheduling scheme. Notice that, when the dedicated WET time is cancelled, \emph{i.e.}, $\tau_0=0$, the NOMA-based scheme may still be effective as users can be charged during the time $\tau_i, \forall i\in \mathcal{N}_0 $. 


During the dedicated WET time slot, only the PS transmits the energy signal to charge the users, thus the harvested energy at the user $n$ is given by
\begin{align}
e_n^0(t)=\frac{\tau_0 \eta P_s L_0 d_0^{\alpha} |g_n(t)|^2}{R_n^\alpha}, \forall n \in \mathcal{N},
\end{align}
where $\eta$ denotes the energy conversion efficiency.

During the time slot that the user $i\in\mathcal{N}_0$  sends information to the AP, the PS still charges the other users except the transmitting user, the received signal at an user $n$ ($n\neq i$) is given as
\begin{equation}
\begin{aligned}
y_n^{i}(t)\!=\!\!\sqrt{\frac{P_sL_0d_0^{\alpha}}{R_n^\alpha}}g_n(t) s(t)\!+\!\!\!\sqrt{P_{i}(t)}h_{in}(t)s_{i}(t)\!+\!z_n(t),
\end{aligned}
\end{equation}%
where $s(t)$ and $s_{i}(t)$ represent the signal transmitted by the PS and the user $i$ in $t$-th time block, $h_{in}(t)$ denotes the channel fading between the user $n$ and $i$, and $z_n(t)$ is the noise.
In a WPCN, the transmission power of user $i$ is much smaller than that of the PS, and the second term is much smaller than the first term.
Then the second term in (2) can be ignored, and the harvested energy at the user $n$ can be written by
\begin{align}
e_n^{i}(t)=\frac{\tau_{i} \eta P_sL_0d_0^{\alpha}  |g_n(t)|^2}{R_n^\alpha}, \ \forall n \in \mathcal{N} \ \mathrm{and} \ n\neq i.
\end{align}
Thus, the total amount of harvested energy by the $n$-th user in the whole time block is given as
\begin{equation}
\begin{aligned}
e_n(t)=\sum_{i\in\{0\}\bigcup\mathcal{N}_0, i\neq n}\frac{\tau_{i} \eta P_sL_0d_0^{\alpha}  |g_n(t)|^2}{R_n^\alpha}.
\end{aligned}
\end{equation}%
In order to guarantee the energy sustainability at the users, the amount of consumed energy in one time block should not exceed that of harvested energy in the current time block. 
Given the transmission time $\tau_n$, the transmission power of the user $n$ is written as
\begin{align}
P_n(t)=\frac{e_n(t)}{\tau_n}=\frac{\sum_{i\in\{0\}\bigcup\mathcal{N}_0,i\neq n}\tau_i\eta P_sL_0d_0^{\alpha}  |g_n(t)|^2}{\tau_nR_n^\alpha}.
\end{align}%
Then, when the user $n$ sends its information to the destination, the received signal at the AP is
\begin{equation}
\begin{aligned}
& y^n(t)=\\
& \ \!\!\!
\begin{cases}
\sqrt{\!\frac{P_n(t)L_0d_0^{\alpha} }{(r_d-R_n)^\alpha}}h_{n}(t)s_n(t)\!+\!\!\!\sqrt{P_s}f(t) s(t)\!+\!z(t), \mathrm{if} \ n \in \mathcal{N}_0, \\
\sqrt{\!\frac{P_n(t)L_0d_0^{\alpha} }{(r_d-R_n)^\alpha}}h_{n}(t)s_n(t)\!+\!z(t), \qquad \qquad \quad \ \  \mathrm{otherwise},
\end{cases}
\end{aligned}
\end{equation}%
where $f(t)$ denotes the channel fading from the PS to the AP after interference cancellation. In the case of $n \in \mathcal{N}_0$, the first and second terms are the expected signal and interference respectively. The signal-interference-noise-ratio (SINR) of the user $n$ is given as below
\begin{equation}
\begin{aligned}
\label{eq:SNR}
& \gamma_n(t)= \\
& \
\begin{cases}
\frac{\eta P_s L_0^2 d_0^{2\alpha}|g_n(t)|^2|h_n(t)|^2\sum_{i\in\{0\}\bigcup\mathcal{N}_0, i\neq n}\tau_i}{\tau_n(P_s|f(t)|^2+\sigma^2)[R_n(r_d-R_n)]^\alpha}, & \mathrm{if} \ n \in \mathcal{N}_0,\\
\frac{\eta P_s L_0^2 d_0^{2\alpha} |g_n(t)|^2|h_n(t)|^2\sum_{i\in\{0\}\bigcup\mathcal{N}_0}\tau_i}{\tau_n\sigma^2[R_n(r_d-R_n)]^\alpha}, & \mathrm{otherwise}.
\end{cases}
\end{aligned}
\end{equation}%
Thus, the expected achievable rate of the user $n$ is given as
\begin{align}
\label{eq:SumRate}
& r_n^l=\mathbb{E}\big[\tau_n\log_2\big(1+\gamma_n(t)\big)\big],
\end{align}%
where the expectation is taken over the small-scale fading $g_n(t),h_n(t)$ and the transmission distance $R_n$.

\section{Fairness-Aware NOMA-based Scheduling Scheme for Wireless Powered Communication Network}
In this section, we propose a fairness-aware NOMA-based scheduling scheme to mitigate the throughput unfairness resulting from the \emph{near-far} problem in a WPCN.
Specifically, we first sort all users according to the channel conditions, and divide them into two groups, \emph{i.e.}, the interference group and the non-interference group. WET and WIT can be scheduled concurrently for NOMA transmissions. To achieve the max-min fairness, we further analyze the achievable rates of the ordered users, and decide the optimal user grouping policy for NOMA transmissions to maximize the minimal rate of users.

This paper aims to unveil the effect of random user deployment on the network fairness. Therefore, we defer the research on the effect of small-scale fading into our future work, and either the constant or the statistical average small-scale fading is assumed in this paper. 
Define the equivalent transmission distance $X_n=R_n(r_d-R_n)$ as the product of the transmission distance for WET and WIT, then the achievable rate of the user $n$ can be rewritten as
\begin{equation}
\begin{aligned}
& r_n^l=
& \
\begin{cases}
\frac{1-\tau_0}{N}\mathbb{E}\big[\log_2\big(1+\frac{(l-1+\kappa)a_n}{bX_n^\alpha}\big)\big]
, & \mathrm{if} \ n \in \mathcal{N}_0,\\
\frac{1-\tau_0}{N}\mathbb{E}\big[\log_2\big(1+\frac{ (l+\kappa)a_n}{X_n^\alpha}\big)\big], & \mathrm{otherwise}.
\end{cases}
\end{aligned}
\end{equation}%
where $a_n=\frac{\eta P_sL_0^2d_0^{2\alpha}|g_n(t)|^2|h_n(t)|^2}{\sigma^2}$, $b=1+\frac{P_s|f(t)|^2}{\sigma^2}$, and $\kappa=\frac{N\tau_0}{1-\tau_0}$.

In a WPCN, the user far from the PS harvests less energy, and the user far from the AP consumes more energy, thus the \emph{near-far} problem causes severe throughput fairness issue.
As an important fairness metric, the max-min fairness to maximize the minimum rate of all users, is investigated.
The user with smaller distance has a larger transmission rate and better capability to combat the interference.
Thus, in order to achieve a better fairness,
we should improve the performance of users with poor channel conditions by sacrificing that of users with good channel conditions.
To achieve this, we first sort all users by the generalized equivalent distance in an ascend order, namely $X_{(1)}\leq X_{(2)}\leq \cdots \leq X_{(N)}$, where $X_{(n)}$ is n-th smallest observation among $X_1 \cdots X_N$, namely the equivalent transmission distance of the $n$-th ordered user. Then, we divide users into two groups,
the interference group $\mathcal{N}_0$ with first $l$ ordered users, and the non-interference group $\mathcal{N}_1$ with the other $N-l$ users.
During the data transmissions of the users in the interference group, the PS concurrently transmits the energy signal to the other users.
With this policy, the transmission rates of the users in the interference group decrease due to interference of NOMA transmission, while those of the users in the non-interference group are enhanced due to more harvested energy. Thus the fairness can be improved, and we refer it as a fairness-aware NOMA-based scheduling scheme.
To this end, the achievable rate of $n$-th ordered user can be rewritten as
\begin{equation}
\begin{aligned}
& r_{(n)}^l\!\!=
& \ \!\!\!\!\!\!\!\!
\begin{cases}
\frac{1-\tau_0}{N}\mathbb{E}\big[\log_2\big(1+\frac{(l-1+\kappa)a_{(n)}}{bX_{(n)}^\alpha}\big)\big]
, & \mathrm{if} \ n \leq l,\\
\frac{1-\tau_0}{N}\mathbb{E}\big[\log_2\big(1+\frac{ (l+\kappa)a_{(n)}}{X_{(n)}^\alpha}\big)\big], & \mathrm{otherwise}.
\end{cases}
\end{aligned}
\end{equation}%

To achieve the max-min fairness, it's critical to decide the set of users in each group. As users are ordered based on $X_n$, the user with the largest index in each group represents the greatest distance and the smallest transmission rate, and to decide the users set is equivalent to determine the optimal number of interfered users $l^*$ as following,
\begin{align}
l^*=\arg\max_l\min\{r_{(l)}^l,r_{(N)}^l\}
\end{align}
where $r_{(l)}^l$ and $r_{(N)}^l$ are the minimum rate of the users in $\mathcal{N}_0$ and $\mathcal{N}_1$, respectively.

We first present a result of order statistics \cite{yang2011order} that is used for the performance analysis of our proposed scheme.
\begin{proposition}
Let $Z_1,Z_2,\dots, Z_N$ be $N$ independent and identically distributed (i.i.d) random variables with common pdf $f_Z(z)$ and CDF $F_Z(z)$. For any $n\in\{1,2,\dots,N\}$, let $Z_{(n)}$ be the n-th order statistics of $Z_1,Z_2,\dots, Z_N$, \emph{i.e.}, $Z_{(n)}$ is the n-th smallest random variable among $Z_1,\dots, Z_N$. Then, the CDF and the pdf of $Z_{(n)}$ are given by 
\begin{equation}
\begin{aligned}
F_{Z_{(n)}}(z)& =\sum_{i=n}^N C_N^i (F_Z(z))^i (1-F_Z(z))^{N-i}\\
& =I_{F_Z(z)}(n,N-n+1),
\end{aligned}
\end{equation}%
\begin{equation}
\begin{aligned}
f_{Z_{(n)}}(z)=&\frac{N!}{(n-1)!(N-n)!}\\
& \cdot(F_Z(z))^{n-1} (1-F_Z(z))^{N-n}f_Z(z),
\end{aligned}
\end{equation}%
where $I_{\cdot}(\cdot,\cdot)$ is the generalized incomplete beta function.
\end{proposition}

With order statistics, we analyze the asymptotic achievable rate of each user in the case that the distance between the PS and AP is much larger than the WET radius, namely, $r_d\gg 2r_e$,  and the SNR of the user is high, \emph{e.g.}, $\gamma_n(t)\gg 1$.
\begin{lemma}
\label{theorem:asymptotic}
In the case of $r_d\gg 2r_e$ and $\gamma_n(t)\gg 1$, we have the approximate expressions $X_n=R_n(r_d-R_n)\approx r_d R_n$ and $r_n^l\approx\frac{1-\tau_0}{N}\mathbb{E}[\log_2(\gamma_n(t))]$, thus the achievable rate of the $n$-th ordered user is given by
\begin{equation}
\begin{aligned}
r_{(n)}^l\approx\frac{1-\tau_0}{N\ln 2}
\begin{cases}
\mathbb{E}\big[\ln\big(\frac{a_{(n)}}{b}\big)\big]+\ln(l-1+\kappa)-\alpha\big[\ln(r_er_d) \\
\quad +\psi(n)-\psi(N+1)\big], \ \mathrm{if} \ n \leq l, \\
\mathbb{E}[\ln(a_{(n)})]+\ln(l+\kappa)-\alpha\big[\ln(r_er_d) \\
\quad +\psi(n)-\psi(N+1)\big], \ \mathrm{otherwise},
\end{cases}
\end{aligned}
\end{equation}%
where $\psi(n)$ is Euler Psi function. Meanwhile, we can derive that $r_{(n)}^l$ monotonically increases and decreases with respect to $l$ and $n$ respectively.
\end{lemma}

\textbf{Proof:}
We only present the proof of (14) for the case that the $n$-th order user is in the interference group $\mathcal{N}_0$, and the other case can be verified using similar arguments.
In the case of $r_d\gg 2r_e$ and $\gamma_n(t)\gg 1$, we have $X_n\approx r_d R_n$, and the order of $X_n$ is same with that of $R_n$, namely $X_{(n)}\approx r_d R_{(n)}$, where $R_{(n)}$ is is n-th smallest observation among $R_1 \cdots R_N$. By (9), the achievable rate of $n$-th ordered user can be rewritten as
\begin{equation}
\begin{aligned}
r_{(n)}^l & = \frac{1-\tau_0}{N\ln 2}\big\{\mathbb{E}\big[\ln\big(\frac{a_{(n)}}{b}\big)\big]+\ln(l-1+\kappa)\\
& -\alpha\big[\ln(r_d)+\mathbb{E}[\ln(R_{(n)})]\big]\big\}.
\end{aligned}
\end{equation}%
The second term is dependent on the transmission power, and the larger $l$ represents the greater amount of harvested energy and higher transmission power. The last two terms are associated with the transmission distance. In the case of $r_d\gg r_e$, the transmission distance from the user to the AP approximately equals the constant $r_d$, and the charging distance from the PS to users becomes the dominated factor.

As users are randomly deployed between the PS and the AP, so the pdf and CDF of the distance between the user and PS $R_n$ are $f_R(r)=\frac{1}{r_e}$ and $F_R(r)=\frac{r}{r_e}$, respectively. By Proposition 1, the pdf of $n$-th order statistics $R_{(n)}$ is given by
\begin{equation}
\begin{aligned}
f_{R_{(n)}}(r)=&\frac{N!}{(n-1)!(N-n)!r_e^{N}}r^{n-1}(r_e-r)^{N-n}.
\end{aligned}
\end{equation}%
Thus, we can derive $\mathbb{E}[\ln(R_{(n)})]$ as
\begin{equation}
\begin{aligned}
& \mathbb{E}[\ln(R_{(n)})] \\
&  \quad = \int_{0}^1 \frac{N!r^{n-1}(1-r)^{N-n}}{(n-1)!(N-n)!}[\ln(r_e)+\ln(r)]dr \\
&  \quad =\ln(r_e)\!+\!\frac{N!B(n,N\!-\!n\!+\!1)}{(n\!-\!1)!(N\!-\!n)!}[\psi(n)\!-\!\psi(N\!+\!1)]\\
& \quad  =\ln(r_e)+\psi(n)-\psi(N+1),
\end{aligned}
\end{equation}%
where $B(n,N-n+1)$ is a beta function and the second equality is derived by the equation index Eq.4.253.1 in \cite{gradshteyn2014table}. Plugging (17) into (15) completes the proof of (14) for the case $n\leq l$. Meanwhile, since $\ln(l-1+\kappa)$ and $\psi(n)$ are the increasing functions with respect to $l$ and $n$ respectively, $r_{(n)}^l$ monotonically increases and decreases with $l$ and $n$ respectively.

Based on the \emph{Lemma} 1, we analyze the monotonic characteristic of $r_{(l)}^l$ and $r_{(N)}^l$, and derive the optimal number of interfered users $l^*$ to achieve the max-min fairness in the following theorem.
\begin{theorem}
In the case of $r_d\gg 2r_e$ and $\gamma_n(t)\gg 1$, the minimum rate $r_{(N)}^l $ of the users in the non-interference group $\mathcal{N}_1$ is monotone increasing with respect to $l$, while the minimum rate $r_{(l)}^l$ of the users in $\mathcal{N}_0$ is monotone decreasing with $l$. Therefore, the optimal $l^*$ to achieve the max-min fairness is determined by bi-section method.
\end{theorem}

\textbf{Proof:} According to the \emph{Lemma} 1, the minimum rate $r_{(l)}^l$ of the user in the interference group is
\begin{equation}
\begin{aligned}
r_{(l)}^l & =\frac{1-\tau_0}{N\ln 2}\Big\{
\mathbb{E}\Big[\ln\Big(\frac{a_{(l)}}{b}\Big)\Big]+\ln(l-1+\kappa)\\
& -\alpha\big[\ln(r_er_d) +\psi(l)-\psi(N+1)\big]\Big\}.
\end{aligned}
\end{equation}%
Thus, when the number of interfered users increases from $l$ to $l+1$, the increment of the minimum rate of users in $\mathcal{N}_0$ can be written as
\begin{equation}
\begin{aligned}
r_{(l+1)}^{l+1}-r_{(l)}^l & =\frac{1-\tau_0}{N\ln 2}\Big\{
\ln\Big(1+\frac{1}{l-1+\kappa}\Big)\\
& -\alpha\big[\psi(l+1)-\psi(l)\big]\Big\}.
\end{aligned}
\end{equation}%
For Euler Psi function, $\psi(l+1)=\psi(l)+\frac{1}{l}$. Meanwhile, $\ln(1+\frac{1}{l+\kappa-1})-\frac{\alpha}{l}$ is a strict monotonic increasing function with respect to $l$ in the case of $\alpha>2$, and $\lim_{z\rightarrow \infty} \ln(1+\frac{1}{l+\kappa-1})-\frac{\alpha}{l}=0$. Thus, we have $r_{(l+1)}^{l+1}-r_{(l)}^l<0$, and the monotonic characteristic of $r_{(l)}^l$ is proofed.
As $r_{(l)}^l$ and $r_{(N)}^l$ monotonically decreases and increases with respect to $l$, respectively, for the max-min fairness problem (11), the bisection-method can be applied to determine the optimal $l^*$.


Usually, the energy transfer radius is much smaller than that of information delivery, therefore we consider $r_d\geq2r_e$ in this paper. The following theorem analyze the achievable rate of each user in a general case.
\begin{theorem}
\label{theorem:rate}
The achievable rate of the $n$-th ordered user that is in the non-interference group $\mathcal{N}_1$, is given by
\begin{equation}
\begin{aligned}
r_{(n)}^l & =\frac{(N-1)!(1-\tau_0)}{(n-1)!(N-n)!}\int_{0}^{1}x^{n-1}(1-x)^{N-n} \\
& \cdot \log_2\Big(1+\frac{(l+\kappa)a_{(n)}r_e^{-2\alpha}}{[x(r_d/r_e-x)]^\alpha}\Big)dx,
\end{aligned}
\end{equation}%
Similarly, the $n$-th ordered user  that is in the interference group $\mathcal{N}_0$, suffers the interference of NOMA transmission, and the corresponding achievable rate is given by
\begin{equation}
\begin{aligned}
r_{(n)}^l& =\frac{(N-1)!(1-\tau_0)}{(n-1)!(N-n)!}\int_{0}^{1}x^{n-1}(1-x)^{N-n} \\
& \cdot \log_2\Big(1+\frac{(l-1+\kappa)a_{(n)}r_e^{-2\alpha}}{b[x(r_d/r_e-x)]^\alpha}\Big)dx.
\end{aligned}
\end{equation}%
\end{theorem}


 \begin{figure*}[t]
      \normalsize
      \centering
      \begin{minipage}[t]{.32\linewidth}
          \centering
          \includegraphics[width=2.5in]{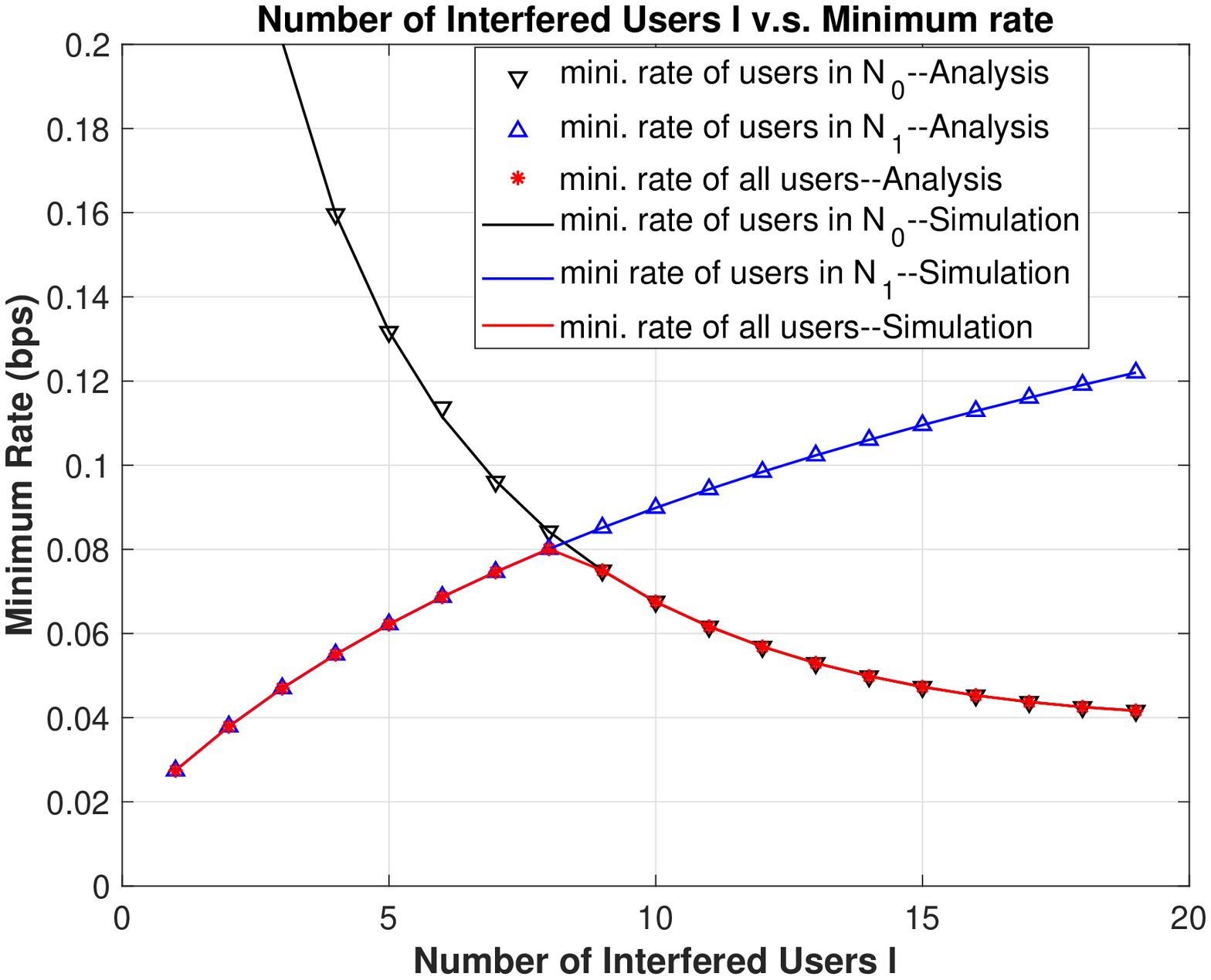}
          \caption{Minimal rate with different number of interfered users.}
          \label{fig:min}
      \end{minipage}
      \begin{minipage}[t]{.32\linewidth}
        \centering
          \includegraphics[width=2.5in]{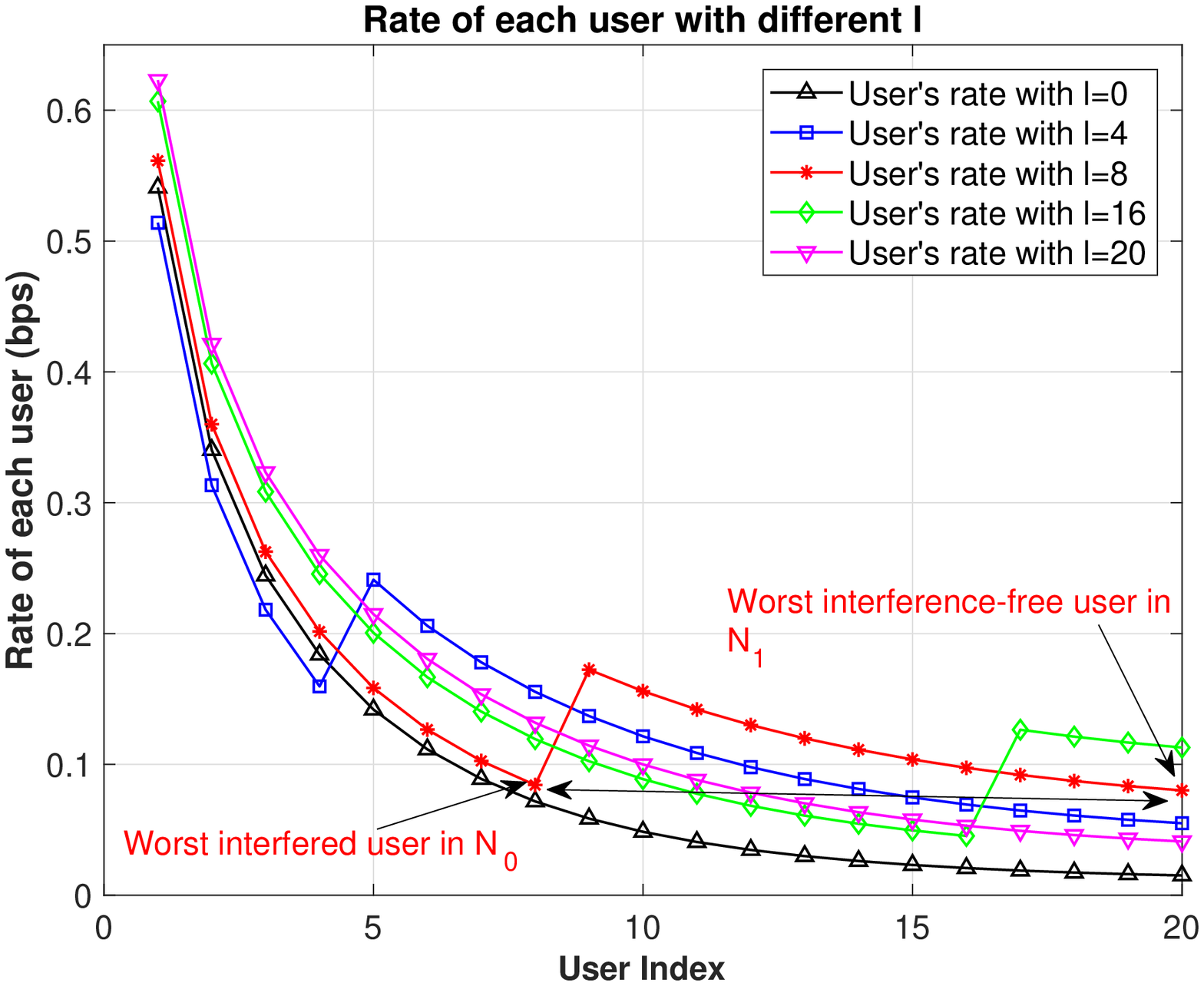}
          \caption{Rate of each user with different number of interfered users.}
          \label{fig:each}
      \end{minipage}
      \begin{minipage}[t]{.32\linewidth}
        \centering
         \includegraphics[width=2.5in]{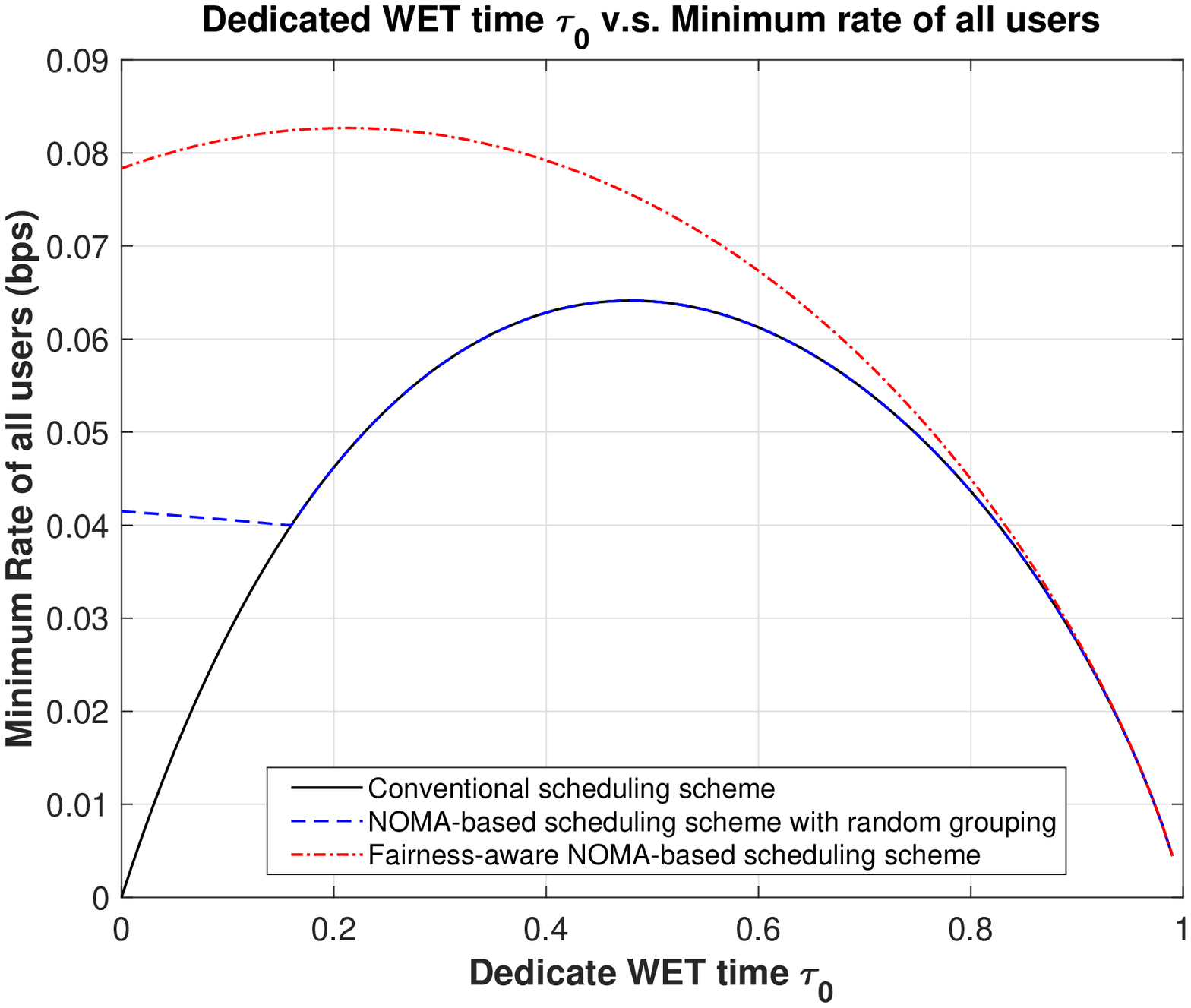}
         \caption{Minimal rate with different dedicated WET time $\tau_0$.}
         \label{fig:tau}
      \end{minipage}
 \end{figure*}%

\textbf{Proof:}
As $r_d\geq 2r_e$, we have $R_n\leq \frac{r_d}{2}$ and $X_n=R_n(r_d-R_n)=\frac{r_d^2}{4}-[\frac{r_d}{2}-R_n]^2$. Thus, the order of $X_n$ is same with that of $R_n$, namely $X_{(n)}=\frac{r_d^2}{4}-[\frac{r_d}{2}-R_{(n)}]^2$.
Given the pdf of the $n$-th order  statistics $R_{(n)}$ in (16), the achievable transmission rate of the $n$-th order user in the non-interference group $\mathcal{N}_1$ can be written as
\begin{equation}
\begin{aligned}
r_{(n)}^l & =
\frac{1-\tau_0}{N}\mathbb{E}\Big[\log_2\Big(1+\frac{(l+\kappa)a_{(n)}}{[R_{(n)}(r_d-R_{(n)})]^\alpha}\Big)\Big]\\
&\ =\frac{(N-1)!(1-\tau_0)}{(n-1)!(N-n)!r_e^N}\int_{0}^{r_e}r^{n-1}(r_e-r)^{N-n}\\
& \ \cdot \log_2\Big(1+\frac{(l+\kappa)a_{(n)}}{[r(r_e-r)]^\alpha}\Big)dr.\\
\end{aligned}
\end{equation}%
By replacing $r$ with $x=\frac{r}{r_e}$, the expression (20) is derived. The rate of the user in the interference group is similarly derived.
Based on the \emph{Theorem} 2, the achievable sum-rate is the sum of the achievable rate of all users, and is calculated by $r^l=\sum_{n=1}^lr_{(n)}^l+\sum_{n=l+1}^Nr_{(n)}^l$.

\section{Performance Evaluation}
In this section, we evaluate the performance of our proposed fairness-aware NOMA-based scheduling scheme, and compare it with the conventional TDMA scheduling scheme and the NOMA-based  scheduling scheme with random grouping policy, where the same time allocation policy is considered.

Unless otherwise stated, we set the number of users $N=20$, the distance from the PS to the AP is $50$ meters, and the WET radius $r_e$ is $20$ meters.
The reference path loss $L_0$ is $0.1$, the path loss exponent $\alpha$ is set to $3$. The noise power $\sigma^2$ is $-70$dBm.
Energy conversion efficiency is $0.5$, and the transmission power at the PS equals $1$, \emph{i.e.}, $P_s = 1$. The strength of interference channel is about $-63$dBm. For the time allocation, we set the dedicated WET time is same with the allocated WIT time for each user, \emph{i.e.}, $\tau_0=\tau_n=\frac{1}{N+1}$.

%

The minimum rate of the users with different numbers of interfered users $l$ is presented in Fig. \ref{fig:min}. As is shown, the minimum rate of the users in the interference group $\mathcal{N}_0$ monotonically decreases with respect to $l$, while that of the users in the non-interference group $\mathcal{N}_1$ monotonically increases. This result is consistent with \emph{Theorem} 1. Therefore, the optimal number of interfered users $l^*$ can be determined by bi-section method, and the optimal $l^*$ is $8$ in this case.
In Fig. \ref{fig:each}, we present the rates of users with different numbers of interfered users $l$. If $l=0$, the fairness-aware NOMA-based scheduling scheme (noted as black line) is degraded to the conventional TDMA scheduling scheme.
It is shown that the user's rate monotonically decreases with respect to the sorting order, and the rate of the user with smallest distance is almost $10$ times of that of the user with largest distance. Thus, the scheme achieves very poor throughput fairness.
Since the non-orthogonal transmission during the data transmissions of the users in $\mathcal{N}_0$, the transmission rates of the interfered users decrease, while the rates of the other users are improved. When $l=8$, the transmission rate of the worst non-interfered user approaches to that of the worst interfered user, and the max-min fairness is achieved. If the number of interfered users is further increased, the achieved rate of the worst user in $\mathcal{N}_0$ will be smaller than that of the worst user in $\mathcal{N}_1$.

 \begin{figure*}[t]
      \normalsize
      \centering
      \begin{minipage}[t]{.32\linewidth}
          \centering
          \includegraphics[width=2.5in]{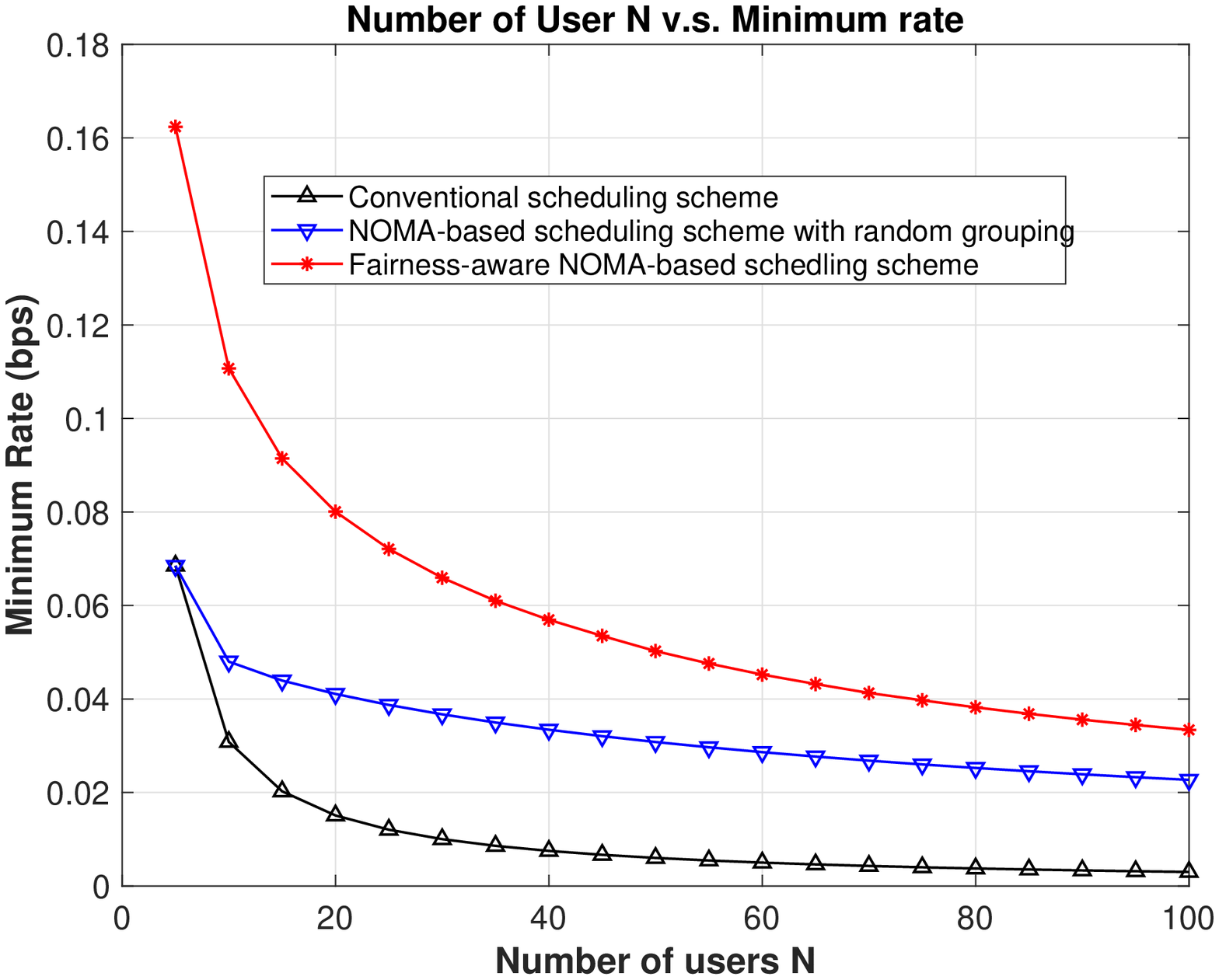}
          \caption{Minimum rate  with different number of users $N$.}
          \label{fig:n_min}
      \end{minipage}
      \begin{minipage}[t]{.32\linewidth}
        \centering
          \includegraphics[width=2.5in]{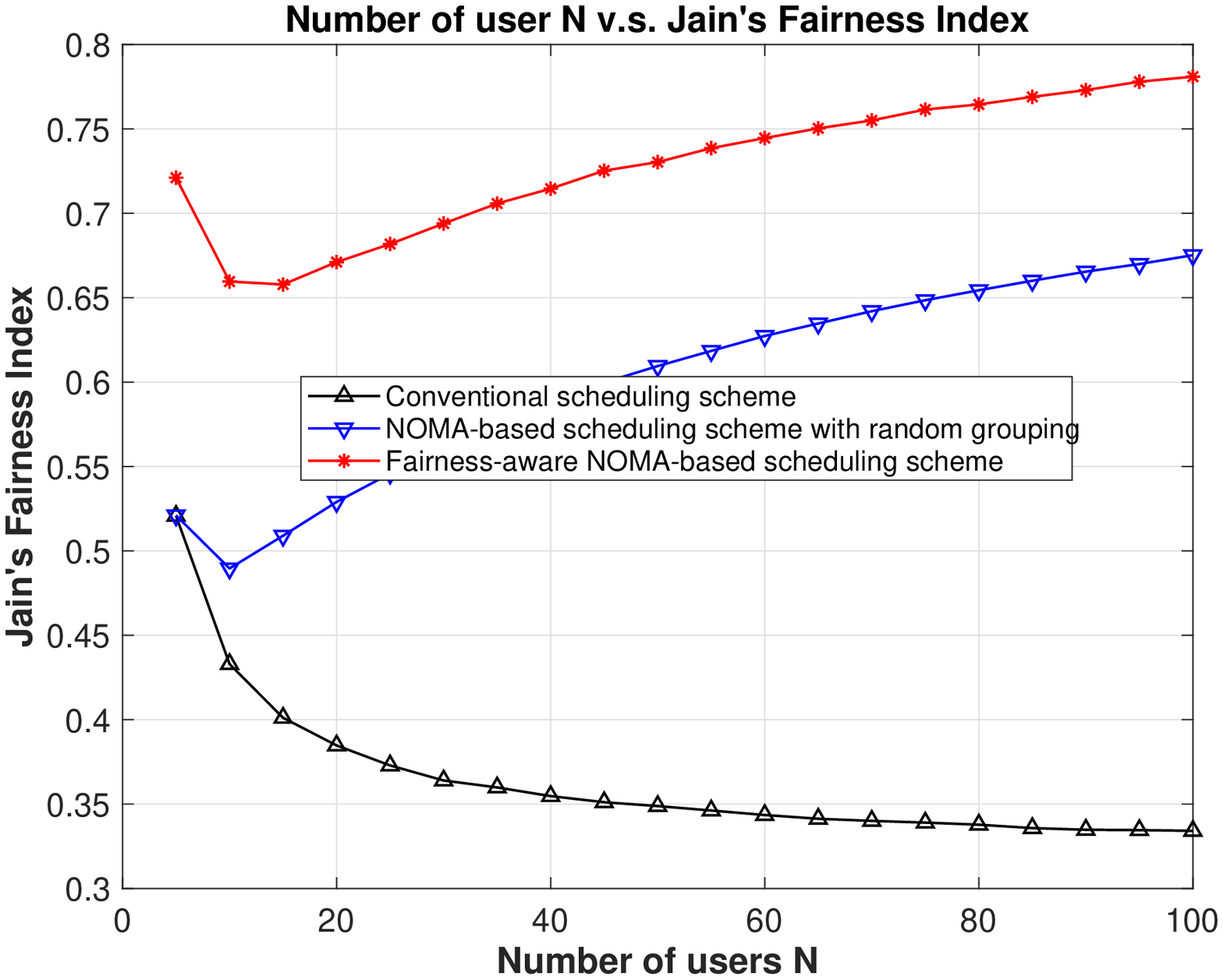}
          \caption{Jain's fairness with different number of users $N$.}
          \label{fig:n_jains}
      \end{minipage}
      \begin{minipage}[t]{.32\linewidth}
        \centering
        \includegraphics[width=2.5in]{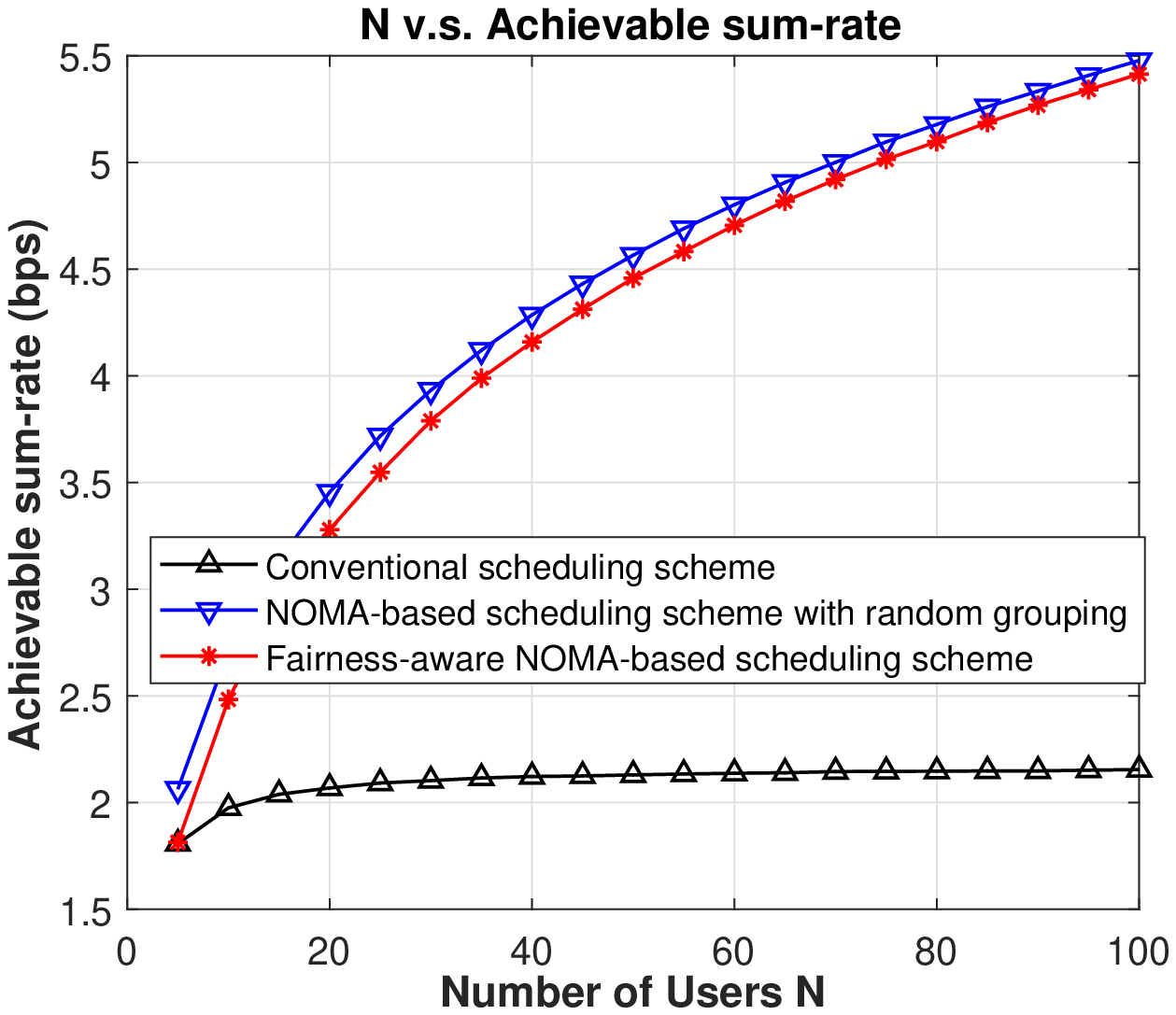}
          \caption{Sum-rate with different number of users $N$.}
          \label{fig:n_throughput}
      \end{minipage}
 \end{figure*}%



The effect of the dedicate WET time $\tau_0$ on the minimum rate is shown in Fig. \ref{fig:tau}. The fairness-aware NOMA-based scheduling scheme achieves the best performance in the terms of max-min fairness.
If $\tau_0=0$, the minimum rate of conventional scheduling scheme is $0$, while that of the NOMA-based scheduling schemes is non-zeros. Therefore, our proposed NOMA-based scheduling schemes are effective even without the dedicated WET time.
When $\tau_0$ is less than $0.16$, compared with conventional scheduling scheme, the NOMA-based scheduling scheme with random grouping achieves a better minimum rate. If $\tau_0> 0.16$, the NOMA-based scheduling scheme with random grouping achieves the same performance with conventional scheduling scheme, which indicates that $l^*$ almost equals zero and NOMA-based scheduling scheme with random grouping is degraded to conventional scheduling scheme.

%



%
%

Fig. \ref{fig:n_min} and Fig. \ref{fig:n_jains} evaluate the max-min fairness and jain's fairness index with varying the number of users $N$ from $5$ to $100$, respectively. Our proposed NOMA-based scheduling schemes significantly improve the network fairness in both terms of max-min fairness and jain's fairness. Meanwhile, the performance of fairness-aware NOMA-based scheduling scheme significantly outperforms that of the NOMA-based scheduling scheme with random grouping, and it is validated that the user grouping policy according to the channel condition is necessary, especially to improve the network fairness.
The achievable sum-rate with different number of users $N$ is presented in Fig. \ref{fig:n_throughput}. It shows that our proposed NOMA-based scheduling schemes could also bring much throughput gain. Compared with the NOMA-based scheduling scheme with random grouping, the fairness-aware NOMA-based scheduling scheme achieves the better fairness at the cost of slightly degraded network throughput. The above results validate that our proposed NOMA-based scheduling schemes could not only improve the network fairness, but also enhance the network throughput.

\section{Conclusion}
This paper has proposed a fairness-aware NOMA-based scheduling scheme for a wireless powered communication network with random user deployment, and analyzed the achieved max-min fairness.
{We have divided users into two groups, \emph{i.e.},  the interference group with good channel conditions and the non-interference group with poor channel conditions. By allowing NOMA transmissions for both WET and WIT, the fairness performance can be improved.}
By applying order statistics theory, we have analyzed the achievable rates of users, and derived the optimal number of interfered users to achieve the max-min fairness.


In the future work, the joint effect of random user deployment and small-scale fading in WPCN will be studied. We will also incorporate the NOMA-based WIT that allows multiple users simultaneously transmit information to the AP, along with the WET for a WPCN. We will further extend the work by considering more network scenarios, such as multiple power stations and access points.



\section*{Acknowledgment}
This research is partially supported by the National Natural Science Foundation of China (NSFC) under grant 61771406, and the National Natural Science Foundation (NSF) under grant ECCS-1554576.

\bibliographystyle{IEEEtran}
\bibliography{IEEEabrv,icc2019}



%
%
%

\end{document}